\documentstyle[psfig,prb,aps]{revtex}
\begin{document}
\draft

\twocolumn[\hsize\textwidth\columnwidth\hsize\csname
@twocolumnfalse\endcsname

\title{
Structures and Stabilities of CaO and MgO Clusters and
Cluster Ions: An alternative interpretation
of the experimental mass spectra.
}
\author{Andr\'es Aguado$^*$ and Jos\'e M. L\'opez}
\address{Departamento de F\'\i sica Te\'orica,
Universidad de Valladolid, Valladolid 47011, Spain}
\maketitle
\begin{abstract}
The structures and relative stabilities of doubly-charged nonstoichiometric
(CaO)$_n$Ca$^{2+}$ (n=1--29) cluster ions and of neutral stoichiometric 
(MgO)$_n$ and (CaO)$_n$ (n=3,6,9,12,15,18) clusters are studied through
{\em ab initio} Perturbed Ion plus polarization calculations. The large
coordination-dependent polarizabilities of oxide anions favor the formation
of surface sites, making the critical cluster size where anions with bulk
coordination first appear larger than that found
in the related case of alkali halides.
Thus, we show that there are substantial structural differences between alkali
halide and alkaline-earth oxide cluster ions, contrary to what is suggested by
the similarities in the experimental mass spectra. An
alternative interpretation of the
magic numbers for the case of oxides is proposed, which involves an explicit
consideration of isomer structures different from the ground states. A
comparison with the previously studied (MgO)$_n$Mg$^{2+}$ cluster ions
shows that the emergence of bulklike structural properties with size is slower
for calcium oxide. Nevertheless, the structures of the doubly charged
clusters are rather similar for the two materials. On the contrary,
the study of the neutrals reveals interesting structural differences between
MgO and CaO, similar to those found in the case of alkali halides.
\end{abstract}

\vskip2pc]

\section{Introduction}

Small clusters are of great interest both to the physical and chemical
communities because of their numerous potential applications (for example, in
nanoelectronics or catalysis), and also because one can gain important insight
into the evolution from atomic and molecular properties to bulk and surface 
properties. To have a knowledge of the structures adopted by the clusters is of
paramount importance, as many interesting cluster properties are largely
determined by them. From the theoretical side, finding the lowest energy
structure for each cluster is a complicated matter, because
the number of isomers increases exponentially with cluster size. Other reasons
are that one has to treat bulklike and surfacelike ions on an equal footing, and
that the number of ions to be explicitely considered in a cluster is larger
than in a bulk or surface study, where symmetry restrictions impose a number
of useful atomic equivalences. From the experimental side, the problem is so
difficult that during the approximately 30 years of intensive cluster research,
the main source of structural information has been theory. Very recently,
experimental techniques like electron diffraction from trapped clusters
\cite{Mai99} or measurements 
of cluster mobilities \cite{Hel91,Jar95,Mai97,Dug97,Hud97}
have been succesfully applied to study the structures of covalent and ionic 
clusters. Photoelectron spectroscopy has also been applied to study 
isomerization transitions in small alkali halide clusters,\cite{Fat96} and
measurements of ionization potential to detect structural transitions in barium
oxide clusters.\cite{Bou98}
At the moment, however, these techniques need parallel theoretical
calculations to make a definite
assignment of the observed diffraction pattern,
mobility or ionization potential to a specific isomer geometry.

A large amount of theoretical work has been devoted to metallic,
semiconductor and noble gas clusters. The work on ionic materials has been
centered mostly in the family of alkali halides, and studies of metal oxide
clusters have been comparatively scarce, despite their importance in many
branches of surface physics, like heterogeneous catalysis or corrotion. 
Saunders\cite{Sau88,Sau89} reported mass spectra and collision induced 
fragmentation data for stoichiometric (MgO)$_n^+$ and (CaO)$_n^+$ cluster ions,
Martin and Bergmann\cite{Mar89} published mass spectra of (CaO)$_n$Ca$^{2+}$
cluster ions, and Ziemann and Castleman
\cite{Zie91a,Zie91b,Zie91c,Zie92} performed experimental measurements of several
singly- and doubly-ionized cluster ions of MgO and CaO by using
laser-ionization time-of-flight mass spectrometry.
Theoretical calculations have been performed at different levels of accuracy:
simple ionic models based on phenomenological pair potentials
were used by Ziemann and Castleman \cite{Zie91b,Zie91c}
to explain the global trends found in their experiments; Wilson \cite{Wil97} has studied neutral (MgO)$_n$ (n$\le$30) clusters by using a compressible-ion
model \cite{Wil96a} that includes coordination-dependent oxide
polarizabilities; \cite{Fow88,Jem98} semiempirical
tight-binding calculations for MgO clusters
were reported by Moukouri and Noguera;
\cite{Mou92,Mou93} finally, {\em ab initio} calculations on stoichiometric
MgO clusters have been presented recently by Recio {\em et al.},
\cite{Rec93a,Rec93b} Malliavin and Coudray, \cite{Mal97} Li {\em et al,}
\cite{Li97} and de la Puente
{\em et al,} \cite{Pue97}
and calculations on stoichiometric (Li$_2$O)$_n$
clusters have been reported by Finocchi and Noguera.\cite{Fin96}
Regarding the nonstoichiometric cluster ions,
Aguado {\em et al.}\cite{Agu99} have studied the structures and stabilities of
(MgO)$_n$Mg$^{2+}$. 

Trying to find an interpretation of the obtained mass spectra,
Ziemann and Castleman\cite{Zie91c} performed some simple pair potential
calculations of the structures of (MgO)$_n$Mg$^{2+}$ cluster ions by using a
rigid ion model. The conclusion of those calculations was that the magic numbers
can be explained in terms of highly compact structures that can only be
obtained for certain cluster sizes, an interpretation very similar to that found
in the closely related case of alkali halides.\cite{Agu97a,Agu97b,Agu98}
In our previous work,\cite{Agu99} we showed that the structures of 
(MgO)$_n$Mg$^{2+}$ cluster
ions were quite different from those of alkali halides. Specifically, the
influence of the large and coordination dependent polarizabilities of oxide
anions (not included in the rigid ion model) favors the formation of surface
oxide sites, and thus structures with bulk oxide anions (coordination 6) are
not energetically competitive until large values of the number of molecules n
are attained.
For example, a highly compact 3$\times$3$\times$3 cube (where the notation
denotes the number of atoms along three perpendicular edges) is not the ground
state of (MgO)$_{13}$Mg$^{2+}$. Nevertheless, the agreement between the
magic numbers obtained through an examination of the stabilities of the clusters
against the loss of an MgO molecule and the experimental ones is complete. It is
just the interpretation of them in terms of structures that is different, that
is, model-dependent. It is interesting to study a similar system like calcium
oxide in order to assess whether those trends are a general feature of
alkaline-earth oxide clusters or not. Moreover, the experiments of Saunders
\cite{Sau88,Sau89} suggest interesting structural differences between both
materials, as the main fragments observed after collisions with inert gas ions
were (MgO)$_3$ in one case and (CaO)$_2$ in the other, and the mass spectra
of Ziemann and Castleman\cite{Zie91c,Zie92} show different stabilities in the
small size regime (magic
numbers at n=5,8,11 for (MgO)$_n$Mg$^{2+}$ and at n=5,7,9,11 for 
(CaO)$_n$Ca$^{2+}$), providing
further motivation for our study. From the theoretical point of view, Ca$^{2+}$
is larger than Mg$^{2+}$, so we can expect ionic size packing effects to play
an important role in determining structural differences. Besides, Ca$^{2+}$ has
a polarizability approximately 6 times larger than Mg$^{2+}$, and the 
polarizabilities of the oxide anions are also larger in CaO because the bonding 
is weaker than in MgO.

In this work we present the results of an extensive and systematic
study of (CaO)$_n$Ca$^{2+}$ cluster ions with n up to 29, and of (MgO)$_n$ and
(CaO)$_n$ neutral stoichiometric clusters with n=3,6,9,12,15,18. The rest of the
paper is organized as follows: in Section II we give a brief resume of the
theoretical model employed, as full an exposition has been reported already in
previous works.\cite{Agu97a} The results are presented in Section III, and
the main conclusions to be extracted from our study in Section IV.

\section{The aiPI model and polarization corrections}

The theoretical foundation of the {\em ab initio} perturbed ion model\cite{Lua90}
lies in the theory of electronic separability,\cite{McW94,Fra92}
and its practical implementation in
the Hartree-Fock (HF) version of the theory of electronic
separability.\cite{Huz71,Huz73} Very briefly,
the HF equations of the
cluster are solved stepwise, by breaking the cluster wave
function into local group functions (ionic in nature in our
case). In each iteration, the total energy is minimized with respect to
variations of the electron density localized in a given ion, with the electron
densities of the other ions kept frozen. 
In the subsequent iterations each frozen
ion assumes the role of nonfrozen ion.
When the self-consistent process finishes,\cite{Agu97a}
the outputs are the total
cluster energy and a set of localized wave functions, one for
each geometrically nonequivalent ion of the cluster.
These localized cluster-consistent ionic wave
functions are then used to estimate the intraatomic
correlation energy correction through Clementi's
Coulomb-Hartree-Fock method.\cite{Cle65,Cha89}
The large multi-zeta basis sets of Clementi and
Roetti\cite{Cle74} are used for the description of the ions. 
At this respect, our
optimizations have been performed using basis sets (5s4p) for $Mg^{2+}$
and (5s5p) for $O^{2-}$, respectively. Inclusion of diffuse basis functions
has been checked and shown unnecessary.
One important advantage coming from the localized nature of
the model is the linear scaling of the computational effort with the number of
atoms in the cluster. This has allowed us to study clusters with as many
as 59 atoms at a reasonable computational cost.

In our previous work on alkaline-earth oxide clusters,\cite{Agu99} we concluded
that the aiPI model is equivalent to a first-principles version
of the semiempirical breathing shell model.\cite{Mat98} The binding energy of
the cluster can be written as a sum of deformation and interaction terms
\begin{equation}
E_{bind} = \sum_RE_{bind}^R = \sum_R(E_{def}^R + \frac{1}{2}E_{int}^R).
\end{equation}
where the sum runs over all ions in the cluster. The interaction energy term is of the form
\begin{equation}
E_{int}^R = \sum_{S \neq R}E_{int}^{RS} = \sum_{S \neq R}
(E_{class}^{RS} + E_{nc}^{RS} + E_X^{RS} + E_{overlap}^{RS}),
\end{equation}
where the different energy contributions are: the classical electrostatic
interaction energy between point-like ions; the correction to this energy
due to the finite extension of the ionic wave functions; the exchange
interaction energy between the electrons of ion R and those of the other ions
in the cluster; and the overlap repulsive energy contribution.\cite{Fra92}
The deformation energy term $E_{def}^R$ is the self-energy of the ion R.
It is an intrinsically
quantum-mechanical many-body term that accounts for the energy change
associated to the compression of the ionic wave functions upon cluster
formation, and incorporates the correlation contribution to the binding energy.
As the model assumes, for computational simplicity, that the ion densities
have spherical symmetry, the only relevant terms that are lacking from the
{\em ab initio} description are the polarization terms. In a polarizable
point-ion approximation, the polarization contribution to the deformation and
interaction energies is
\begin{eqnarray}
E_{int}^{RS,pol} = -\frac{q_R(\vec\mu_S\vec r_{RS})}{r_{RS}^3}
                   -\frac{q_S(\vec\mu_R\vec r_{RS})}{r_{RS}^3} \nonumber \\
        -3\frac{(\vec\mu_R\vec r_{RS})(\vec\mu_S\vec r_{RS})}{r_{RS}^5}
        + \frac{(\vec\mu_R\vec\mu_S)}{r_{RS}^3},
\end{eqnarray}
\begin{equation}
E_{def}^{R,pol} = \frac{\mu_R^2}{2\alpha_R},
\end{equation}
where $\alpha_R$ is the polarizability of the ion R, and $\vec\mu_R$ the dipole
moment induced on ion R. The new terms added to the interaction energy are
the monopole-dipole and dipole-dipole interaction energy terms. The term added
to the deformation energy represents the energy cost of deforming the charge
density of the ion to create the dipole moment. The point-ion approximation
provides just the asymptotic part of the polarization interaction energy, that
is, it is exact only for large ionic separations. As soon as the ions begin
to overlap, there is an important short-range contribution to the induced dipole
moments,\cite{Szi50,Fow84,Mad96,Row98,Jem99} which is of opposite sign to the asymptotic
limit for anions and may in some specific cases reverse the sign of the
asymptotic value. These effects can be easily acomodated in the formalism by
substituting the asymptotic value of the induced dipole moments by the
following expression:
\begin{equation}
\mu_{\alpha}^{total.R} = \mu_{\alpha}^{asymp,R} + \mu_{\alpha}^{sr,R},
\end{equation}
with
\begin{equation}
\mu_{\alpha}^{asymp,R}=\alpha^R\sum_{S\neq R}\frac{r_{RS,\alpha}}{r_{RS}^3}q^R,
\end{equation}
\begin{equation}
\mu_{\alpha}^{sr,R}=\alpha^R\sum_{S\neq R}\frac{r_{RS,\alpha}}{r_{RS}^3}f(r_{RS}),
\end{equation}
where $sr$ stands for ``short-range'' and $\mu_{\alpha}^{total.R}$ is the
$\alpha$ component of the dipole moment vector induced on ion R. The physics
behind the short-range polarization correction has been explained in Ref.
\onlinecite{Szi50,Fow84}, and is associated to the finite extents of the electron
densities of the anions and cations. Then, $f$ is a short-range function that
switchs on as the cation-anion overlap becomes appreciable. Madden and
coworkers\cite{Jem99} have employed the Tang and Toennies dispersion damping
function\cite{Tan84} as a suitable form for $f$:
\begin{equation}
f(r_{RS}) = -c\sum_{k=0}^{k_{max}}\frac{b^k}{k!}e^{-br_{RS}}.
\end{equation}
This is a smoothed step function passing from zero for large $r$ to $-c$ for
r=0. The range of $r$ values at which $f$ becomes significantly different from
zero is primarily determined by the range parameter $b$.

We have included the polarization terms in the energy calculation with this
parameterised method, that calculates the induced dipole moments from eq. (5)
and the correction to the deformation and interaction energies from eqs. (3) and
(4), respectively. The ``enlarged'' aiPI+polarization model thus obtained
accounts for all the relevant physical interactions. The relaxation of the
assumption of spherical symmetry being computationally expensive, 
the price to be paid
is the inclusion in the model of a set of parameters, namely, the 
polarizabilities $\alpha_R$ and the range parameter $b$. Appropriate values 
for the other two constants $c$ and 
$k_{max}$ can be taken equal to the bulk values ($c$=-3 and $k_{max}$=4).
\cite{Jem99} 
Given the meaning of the range parameter $b$, inversely related
to half the interionic distance between first neighbors, one might expect
different values of $b$ for clusters as compared to bulk materials if the
interionic distances are substantially different. As a matter of fact, the
evolution of those distances with cluster size is not too complicated in the
case of ionic clusters.\cite{Pue97,Agu97a,Agu97b,Agu98} Specifically, the average
interionic distance $d$ initially increases quite abruptly with the number of 
molecules n, and then slowly approaches the bulk limit. As a consequence of
this behaviour, we will see that the bulk value ($b$=0.75 a.u.)\cite{Jem99} is
appropriate for all (CaO)$_n$Ca$^{2+}$ clusters with $n\ge 4$. Different 
values of $b$ are needed just for $n<$4 to avoid overpolarization problems.\cite{Wil97}
Regarding the polarizabilities, oxide anions have the interesting property of 
showing strongly coordination-dependent values. In fact, the O$^{2-}$ anion
does not exist as a free ion, which is equivalent to an infinite polarizability;
in the solid phase it is stabilized by the crystal environment and has a
finite material-dependent polarizability. Wilson\cite{Wil97} has interpolated
between those two limits and gives values for the coordination-dependent
values of $\alpha(O^{2-})$ in MgO. We have assumed that the ratio of the
bulk oxide polarizabilities for MgO and CaO 
($\alpha^{bulk}(O^{2-}:CaO)$/$\alpha^{bulk}(O^{2-}:MgO)$=1.469)\cite{Jem99} is
independent of the oxide coordination, and have deduced the $\alpha$ values for
CaO from those of MgO. This procedure is justified because the cation size does
not change appreciably with coordination number. For the calcium cation we
take the bulk polarizability (3.193 a.u.)\cite{Fow85}

We close this section with a consideration of several criticisms that could be
raised against (and of the advantages of) the employed methodology. We have
chosen a mixed {\em ab initio}/semiempirical energy model in order to obtain a
good compromise between computational efficiency and accuracy. All the
relevant energy terms excluding polarization are described with an {\em ab
initio} methodology. To include polarization, we have used an accurate model,
\cite{Jem99} where the parameters have been fitted by a comparison to
{\em ab initio} calculations.\cite{Jem98} 
Special care has been devoted to the
separation of all the independent physical factors that influence a given
quantity, thus avoiding a mixing of different effects in a single parameter and
enhancing the transferability of the model. The good parameterisation is
reflected in the fact that parameters can be transfered between closely related
systems (like, for example, different metal oxides) by simple scaling arguments
involving ionic radii.\cite{Row99} Thus, we think that the reliability of our
calculations is reduced just a little compared to full {\em ab initio}
methodologies. To support this expectation, we made a comparison with DMOL
calculations performed on neutral (CaO)$_n$ clusters by Malliavin and Coudray.
\cite{Mal97} All the interionic distances were in agreement to their
calculations up to differences of 4 \%. The energetic ordering of the isomers,
as well as the specific energy differences, are reproduced with a maximum error
of 5 \%. We believe that this is a very reasonable agreement, even more if we
realize that we are neglecting dispersion interactions, and polarization
interactions beyond the dipolar terms. The solid MgO is excellently described
with the aiPI model (at least in 
its static properties).\cite{Lua90b} The model is
then expected to transfer properly between both limits.
The larger computational simplicity has been exploited to study large cluster 
sizes (up to 59 ions) with full relaxations of the geometries. Moreover, for
each cluster size, a large number of isomers (between 10 and 15) have been
investigated. The generation of the initial cluster geometries was accomplished
by using a pair potential, as we explained in our previous publication.
\cite{Agu99} The optimization of the geometries has been performed by using a
downhill simplex algorithm.\cite{Nel65,Pre91}

\section{Results and Discussion}

\subsection{Structural Trends in (CaO)$_n$Ca$^{2+}$ Cluster Ions}

In Fig.1 we present the optimized aiPI+polarization structures of the ground
state (GS) and lowest lying isomers or (CaO)$_n$Ca$^{2+}$ (n=4--29)
cluster ions. Below each
isomer we show the energy difference (in eV) with respect to the ground state.
For n$<$4, the clusters are not detected in the experiments, probably because
they undergo a Coulomb explosion driven by the excess charge and the small
cluster size, but we are not interested here in this aspect of the experiments.
From n=4 to n=10, there is a predominance of m$\times$2$\times$2 fragments
(m=2--5),
that is, the (CaO)$_2$ subunit appears as the basic building block. The total
number of ions in these nonstoichiometric clusters is an odd number, and thus
those structures are never perfectly compact. There is either an extra cation
added to or a missing anion removed from the perfect structure. Less compact
structures as for example planar fragments are not energetically competitive.
The structures in this size range tend to be elongated as a direct consequence
of the excess cluster charge. When n=9, a 3$\times$3$\times$2+1 fragment is
more stable than that based on the (CaO)$_2$ building block, and n=10 is the
largest cluster size for which a fragment of this kind is the ground state.
Another thing to be pointed out is that in this size range, the extra cation
present in the m$\times$2$\times$2+1 structures induces a larger cluster
distortion than the missing anion in the m$\times$2$\times$2-1 structures. We
will see that this feature has important implications in the stability of the
clusters.

From n=11 to n=15, the dominant fragments are based on m$\times$3$\times$2
units. For n=16 and 17, the most stable isomers are m$\times$4$\times$2
fragments. None of these structures has still developed an anion with full
bulk coordination. In particular, the 3$\times$3$\times$3 isomer for n=13,
which is particularly stable in the case of nonstoichiometric alkali halide
cluster ions,\cite{Agu98} does not even appear in Fig. 1. 
The large coordination-dependent values
of the polarizabilities of the oxide anions favors the formation of surface 
sites, and gives rise to somewhat less compact ground state structures, for 
which the increase in dipolar energy compensate for the decrease in Madelung
energy. The 3$\times$3$\times$3 (CaO)$_{13}$Ca$^{2+}$ is specially unfavored by
the dipolar energy terms because it has a central oxide anion with bulk
coordination (so with a comparatively low polarizability), and another 12
anions with coordination 4. On the contrary, the largest coordination in the
GS structure is five, and some three-coordinated anions (in corner positions)
also appear, inducing a large dipolar energy stabilization. For n=18 and 19
there is a glimpse of a transition to more compact cluster structures. The
important feature of the GS structures of these two cluster sizes,
compared to the 3$\times$3$\times$3 for n=13, is that now there are oxide
anions in corner positions. These make a large contribution to the polarization
energy term, that added to the increased Madelung energy of a compact
fragment, gives a total GS
energy more negative than that of m$\times$3$\times$2 or
m$\times$4$\times$2 structures. Nevertheless, the energy differences between
isomers are small, and 
from n=20 on, ground state isomers without bulk anions
are again obtained (n=24 and 27 are the only relevant exceptions, because the
ground states of n=26 and n=29 can be considered degenerate within the
accuracy of our theoretical model). 

A general feature of (CaO)$_n$Ca$^{2+}$ cluster ions with n$\ge$8
is that a$\times$b$\times$c+1 fragments are specially
stable compared to other isomers whenever they can be formed. In
Table I we show all the fragments of that kind relevant to the cluster
size range considered in this study. Each series has a typical periodicity that
could in principle be reflected in different portions of the mass spectra, given
the high stability of these fragments. Some sizes can be accomodated in several
families, that is, the classification is highly redundant, but useful anyway
to our purposes. If for a given cluster size, a cluster with that formula can
be formed, it is always the ground state structure. If it is possible to build
up two different isomers with that formula (n=12, 18, 24), the more compact
structure is energetically favored. This rule works as long as we do not
consider structures that are not energetically competitive anymore (the
isomer based on the (CaO)$_2$ building block of (CaO)$_{14}$Ca$^{2+}$ is an
example), and can be helpful in guessing specially stable structures for
clusters larger than those studied here. For nearly all those
cluster sizes with no
competitive a$\times$b$\times$c+1 structure, a$\times$b$\times$c-1 fragments
are obtained as the ground state or specially stable isomers (examples are found
for n=5, 7, 11, 19, 23 and 29). The special stability of a$\times$b$\times$c+1
structures is sometimes reflected in high stabilities for the corresponding
a$\times$b$\times$c+3 structures, comparable indeed to the stabilities of
a$\times$b$\times$c-1 fragments; this occurs for n=13, 17, 19 and 26. With the
only exceptions of n=13,14,22,26 and 29, all (CaO)$_n$Ca$^{2+}$ GS structures 
are explained in terms of those three kinds of fragments.

Comparing to the results of our previous paper on (MgO)$_n$Mg$^{2+}$ cluster
ions,\cite{Agu99} we can see that from n=4 to n=20 the GS structures are 
basically the
same in both systems (the only exceptions are n=7 and n=13). Interesting
structural differences between both materials appear in the size range n$>$20.
Specifically, the transition to bulklike structures, containing inner anions 
with bulk coordination, is slower in the case of (CaO)$_n$Ca$^{2+}$. An
analysis of the several energy components shows that the net effect of
polarization is more important in calcium oxide. Although the polarizability
of Ca$^{2+}$ is larger than that of Mg$^{2+}$ and the coordination-dependent
values of $\alpha$(O$^{2-}$) are larger in CaO than in MgO, this is not a
trivial conclusion, because the interatomic distances are also larger in CaO,
and so the electric fields acting on each ion are correspondingly smaller.
If we consider that a highly ionic material is that one for which the Madelung
energy term is almost completely dominant in determining structural and several
other properties, we would conclude that the ionic character of 
(CaO)$_n$Ca$^{2+}$ is smaller that that found for (MgO)$_n$Mg$^{2+}$.
Indeed, being the polarization contribution more important, the structures of
calcium oxide clusters have a larger directionality degree, a feature that is
usually associated to covalency (opposite to the natural tendency of purely
ionic systems to form isotropic structures). However, we think that the term
``covalency'' should be employed just in those situations where charge
transfer between different atomic centers is important. As Madden and coworkers
have discussed,\cite{Mad96} polarization terms in ionic systems are responsible
for a lot of properties traditionally attributed to ``covalency''. One point
that deserves further investigation, however, is whether the directional
properties induced by polarization effects (and reflected in a lower average
coordination) can be responsible for a larger charge transfer between
different centers. This would be reasonable because the saturation of the bonds
is less complete.

\subsection{Relative stabilities and connection to experimental mass spectra}

In the experimental mass spectra,\cite{Mar89,Zie92} the populations
observed for some cluster sizes are enhanced over those of the neighboring
sizes. These ``magic numbers'' are a consequence of the evaporation
events that occur in the cluster beam, mostly after ionization.\cite{Ens83}
A magic cluster of size n has an evaporation energy that is large compared to 
that of the neighboring sizes (n-1) and (n+1). Thus, on the average, clusters
of size n undergo a smaller number of evaporation events and this leads to
the maxima in the mass spectra. As our main concern in this section is to
compare with the experimental mass spectra, we calculate the evaporation
energy as a function of cluster size. To do this, we assume that the dominant
evaporation channel is the loss of a neutral (CaO) molecule, something supported
by the experiments of Ziemann and Castleman.\cite{Zie92} In the size range
n$<$11, some other chanels seem to be opened in the experiments,\cite{Zie92} and
indeed for n$<$4 Coulomb explosion is dominant, that is the reason why we do
not consider clusters with n$<$4. With that assumption, the evaporation energy
of (CaO)$_n$Ca$^{2+}$ reads
\begin{eqnarray}
E_{evap}(n) = E_{cluster}[(CaO)_{n-1}Ca^{2+}] + E(CaO) \nonumber \\
             - E_{cluster}[(CaO)_nCa^{2+}].
\end{eqnarray}
Maxima in the evaporation energy curve do not always coincide with maxima in
the experimental mass spectra.\cite{Agu99} There are two main processes that
contribute to enhance the cluster population for size n: 
a)A small evaporation energy
for size (n+1); b)A large evaporation energy for size n. Thus, a most
convenient quantity to compare with experiment is the second energy difference
\begin{equation}
\Delta_2(n) = E_{evap}(n+1) - E_{evap}(n).
\end{equation}
A negative value of $\Delta_2(n)$ indicates that the n-population increases
by evaporations from the (n+1)--clusters more rapidly than it decays by 
evaporation to the (n-1)--clusters. Specifically, the
specially stable cluster sizes will be reflected as minima in the $\Delta_2(n)$
curve.

Now, the evaporation energy E$_{evap}(n)$ of eq. (9) can be calculated in two 
different ways. In the first one, energy differences are always taken between
the ground state structures of sizes n and (n-1). This procedure, which we
call (by obvious reasons) adiabatic evaporation, reflects the stability
of the clusters in the limit of small energy barriers between isomers
or alternatively of large experimental times of flight. The stabilities
calculated in this way are shown in the upper part of figure 2. 
Magic numbers are found for n=5,8,12,15,18,20,24,27. The only 
a$\times$b$\times$c-1 structure that shows a special stability is that of n=5.
The rest of magic clusters belong to the a$\times$b$\times$c+1 family of
structures. If n is a magic size, and both (n+1) and (n-1) GS structures do not
belong to the a$\times$b$\times$c+1 family, a deep minima is found in the
$\Delta_2(n)$ curve (this happens for n=12 and 18). For the rest of magic sizes,
the (n+1) GS structure has also the formula a$\times$b$\times$c+1, and has a
correspondingly high stability reflected in a negative value of $\Delta_2(n+1)$.
In these cases the stability of size n is just slightly enhanced over that of
size (n+1). One can appreciate the increasing relevance of the Madelung term in
determining the cluster stabilities: when n$<$20, the most stable
a$\times$b$\times$c+1 structures are the less compact ones (n=8 and 15 more
stable than n=9 and 16, respectively); if n$\ge$20, that trend is reversed
(n=20, 24 and 27 more stable than n= 21, 25, and 28, respectively).
The special relevance of a$\times$b$\times$c+1 structures in explaining the
cluster stabilities does not conform to the initial experimental expectations
of high stabilities for a$\times$b$\times$c-1 structures.\cite{Zie92}
Analysing the energy components, we find that the polarization contribution
stabilizes the a$\times$b$\times$c+1 structure more than the corresponding 
a$\times$b$\times$c-1 structure for all values of a,b,c. For the smallest
cluster sizes, however, the extra cation present in a$\times$b$\times$c+1
structures induces a large cluster distortion compared to that induced by the
missing anion in a$\times$b$\times$c-1 structures, and the Madelung
contribution favors these last structures in a larger amount, making them more
stable for some sizes. 

The second kind of calculation of E$_{evap}$(n) proceeds as follows:
we consider the optimized GS structure of (CaO)$_n$Ca$^{2+}$ and identify the
CaO molecule that contributes the least to the cluster binding energy.
Then we remove that molecule and relax the resulting
(CaO)$_{n-1}$Ca$^{2+}$ fragment to the nearest local minimum. 
This process can be termed locally adiabatic because both fragments are allowed
to relax to the local minimum energy configuration after the evaporation. For
some cluster sizes, the fragment of size (n-1) left when a CaO molecule is
removed from (CaO)$_n$Ca$^{2+}$ does not lie on the catchment basin of the
(CaO)$_{n-1}$Ca$^{2+}$ GS isomer, so that the locally adiabatic
evaporation energies
are larger than the energy differences between adjacent ground states minus
E(CaO) in those cases. The locally adiabatic evaporation energies are plotted 
as a function of n in the lower part of Fig. 2. These will reflect the cluster
stabilities in the limit of large energy barriers between isomers or of short
experimental times of flight. Magic numbers are obtained for
n=5,7,9,11,13,16,19,22,25 and 27, in complete agreement with the experiments
of Ziemann and Castleman.\cite{Zie92}

The main message to be extracted from these considerations is that the magic
numbers obtained in the experiments might be dominated by the effects of
kinetic traps occuring in the course of the evaporation process. Since our
calculations are static, 
we can not rigorously assert that this is the only possible
explanation, but a plausibility argument based on a comparison to the closely
related and more thoroughly studied case of alkali halides supports our
expectations. The mobility experiments performed by the group of
Jarrold\cite{Dug97,Hud97} show that the relaxation dynamics to the ground state
structure for sodium chloride clusters involves drift times of almost one
second. The importance of kinetic traps in explaining these interesting results
is shown in the theoretical works of Doye and Wales.\cite{Doy99} Specifically,
these authors show that the potential energy landscape of alkali halide
clusters, calculated by using a phenomenological pair potential to describe the
interactions, is structured in several funnels, separated from each other by
high free-energy barriers. When a cluster evaporates a molecule, it cools in the
process, so trapping kinetic effects are expected whenever parent and product
GS structures belong to different funnels. Given the close similarities
between halide and oxide systems,
one expects similar effects in the evaporation kinetics
of alkaline-earth oxides to be relevant. The main structural differences are
due to the effects of polarization, and these could also affect the
mechanisms of structural transitions. In the case of alkali halides, Doye and
Wales find that a highly cooperative process is energetically less
impeded by energy barriers than sequential ionic diffusion,\cite{Doy99} with
interesting implications for the mechanical properties of these clusters.
Perhaps the same is true for the clusters studied here, but one has to keep in
mind that polarization tends to lower the barriers against 
diffusion,\cite{Wil96} and those effects are more important for oxides.
We think that further
calculations of this kind for oxide clusters would be very interesting. 
Mobility experiments on (MgO)$_n$Mg$^{2+}$ or (CaO)$_n$Ca$^{2+}$ could 
conclusively confirm the structural trends found in the present work.

\subsection{Neutral Stoichiometric (MgO)$_n$ and (CaO)$_n$ clusters}

The experiments performed by Saunders\cite{Sau88,Sau89} show that both
(MgO)$_n^+$ and (CaO)$_n^+$ stoichiometric cluster ions with a number of
molecules n=6,9,12 and 15 are expecially abundant in the mass spectra.
However, when these clusters are allowed to collide with inert gas ions, the
fragmentation channels are different: (MgO)$_3$ fragments are predominantly
observed in one case and (CaO)$_2$ fragments in the other. These results
suggest that the basic cluster building blocks are different for the two
materials, but not so different as to lead to different magic numbers.

We found a similar scenario in the case of alkali halide clusters.\cite{Agu97b}
Specifically, a universal set of magic numbers n=4,6,9,12,... was found for
the whole family of (AX)$_n$ clusters, with A=Li,Na,K,Rb and X=F,Cl,Br,I.
However, the cluster structures were not found to be the same for all the
different materials. When the cation size is much smaller than the anion size
(all lithium halides and sodium iodide),\cite{Agu97a} ground state
structures based on the stacking of hexagonal (AX)$_3$ rings are obtained. For
the rest of materials, the ground state structures are mostly obtained by
stacking of rectangular (or double-chain) (AX)$_3$ planar fragments. This is
just a packing effect: when the ratio of cation to anion size is very small,
anion-anion overlap repulsive interactions are large, forcing an opening of the
(AX)$_3$ rectangular fragments into hexagons. The magic numbers are the same for
both structural families because it is for those cluster sizes that specially
compact structures can be formed. When we studied (AX)$_n$A$^+$ alkali halide
cluster ions,\cite{Agu98} we found that the structures were much more similar
irrespective of packing considerations. The ring structures are not competitive
in this case because it is not possible to build up a perfect hexagonal
fragment with an odd number of ions. On the contrary, perfect cubic structures
can be formed (as for example the 3$\times$3$\times$3 structure for n=13).

From our study on doubly-charged clusters, we have not found important
structural differences between (CaO)$_n$Ca$^{2+}$ and (MgO)$_n$Mg$^{2+}$,
\cite{Agu99} at least in the small size regime. We have performed 
aiPI+polarization calculations on the structures of (MgO)$_n$ and (CaO)$_n$
with n=3,6,9,12,15 and 18.
Specifically, we have considered just those structures based on staking of
hexagonal and rectangular (AO)$_3$ units, and those based on stacking of 
(AO)$_2$ units, with A=Mg or Ca. We find that
the ground state structures of (MgO)$_n$ clusters are based on (MgO)$_3$ units,
while those of (CaO)$_n$ clusters have a rectangular (CaO)$_3$
building block, being this the same packing effect found in the case of alkali
halides. The structure of (CaO)$_6$ could be alternatively viewed as the
stacking of three (CaO)$_2$ units, but for n=9,12,15 and 18, the tubular
shapes obtained by stacking (CaO)$_2$ units are not competitive anymore. 
Were all the ground state structures of (CaO)$_n$ clusters based on the
(CaO)$_2$ building block, we would expect a periodicity of 2 in the magic
numbers observed in the mass spectra.
Saunders shows the collision induced fragmentation spectra of (CaO)$_n$, with
n=4,6,8,\cite{Sau88} which are certainly based on stacking of (CaO)$_2$
units, but does not show those for (CaO)$_9$ or (CaO)$_{12}$,
for example. Our main conclusion is that the special stability
of (CaO)$_n$ clusters is also explained in terms of (CaO)$_3$ units, but with
rectangular instead of hexagonal shape. 
This explains the same periodicities observed
in the magic numbers of both materials.

\section{Summary}

The {\em ab initio} perturbed ion model, supplemented with a parameterised
treatment of dipolar terms, has been employed in order to study
the structural and energetic properties of (CaO)$_n$Ca$^{2+}$ (n=1--29)
cluster ions. Polarization effects favor the formation of surface sites, and
reduce the stability of highly compact structures containing anions with bulk
coordination. Thus, despite many similarities in the experimental mass spectra,
the structures of alkaline-earth oxide and alkali halide cluster ions are shown 
to be different. Most of the lowest energy structures have the formula
a$\times$b$\times$c+1. The structures of (CaO)$_n$Ca$^{2+}$ and 
(MgO)$_n$Mg$^{2+}$ cluster ions are very similar for n$<$20, irrespective of
differences in cationic size and polarization. It is just for n$\ge$20 that
structural differences emerge, showing a slower convergence to bulk properties
for CaO compared to MgO. The analysis of the stabilities suggests that the
experimental mass spectra could be dominated by the effects of kinetic traps.
Specifically, if we consider locally adiabatic evaporation events, complete
agreement is found with the experimental stabilities. The neutral stoichiometric
(MgO)$_n$ and (CaO)$_n$ clusters (n=3,6,9,12,15,18)
show structural differences similar to those observed
in neutral stoichiometric alkali halide clusters: the basic building block is
an (MgO)$_3$ hexagonal fragment in the case of MgO and a (CaO)$_3$ rectangular
(or double-chain) fragment in the case of CaO. This is just a packing effect
due to the larger overlap repulsion between anions when the cation size is very
small. While the structures of (CaO)$_n$ clusters, 
with n=4,6,8 are certainly based on
(CaO)$_2$ units, as suggested by collision-induced fragmentation experiments,
the specially stable (CaO)$_n$ clusters are based on a (CaO)$_3$ unit. This
explains the same periodicity of 3 observed
in the experimental magic numbers of both (MgO)$_3^+$ and
(CaO)$_3^+$ clusters.

$\;$

$\;$



{\bf Captions of Figures and Tables.}

{\bf Figure 1}. Lowest-energy structure and low-lying isomers of
(CaO)$_n$Ca$^{2+}$ cluster ions. Dark balls are Ca$^{2+}$ cations and light
balls are O$^{2-}$ anions. The energy
difference (in eV) with respect to the most stable structure is given below
the corresponding isomers.

{\bf Figure 2}.
Adiabatic (a) and locally adiabatic (b) evaporation energies required to
remove a neutral CaO molecule from
(CaO)$_n$Ca$^{2+}$ cluster ions as a function of n. The local minima in the
evaporation energy curve are shown explicitely.

{\bf Table I}
Possible different a$\times$b$\times$c+1 structures, with their inherent
periodicities. Those cluster sizes n that are actually observed as ground
state structures of (CaO)$_n$Ca$^{2+}$ clusters are written in boldface.

\begin {table}
\begin {center}

\begin {tabular} {|c|c|c|} \hline
 Structure & Periodicity & Cluster size n \\
\hline
m$\times$2$\times$2+1 & 2 & {\bf 8,10},12,... \\
m$\times$3$\times$2+1 & 3 & {\bf 9,12,15},18,{\bf 21},24,27... \\
m$\times$4$\times$2+1 & 4 & {\bf 8,12,16,20},24,{\bf 28},... \\
m$\times$5$\times$2+1 & 5 & {\bf 10,15,20,25},... \\
m$\times$6$\times$2+1 & 6 & 12,18,24,... \\
m$\times$3$\times$3+1 & 9 & {\bf 9,18,27},... \\
m$\times$4$\times$3+1 & 6 & {\bf 12,18,24},... \\
\end {tabular}
\end {center}
\end {table}

\onecolumn[\hsize\textwidth\columnwidth\hsize\csname
@onecolumnfalse\endcsname

\begin{figure}
\psfig{figure=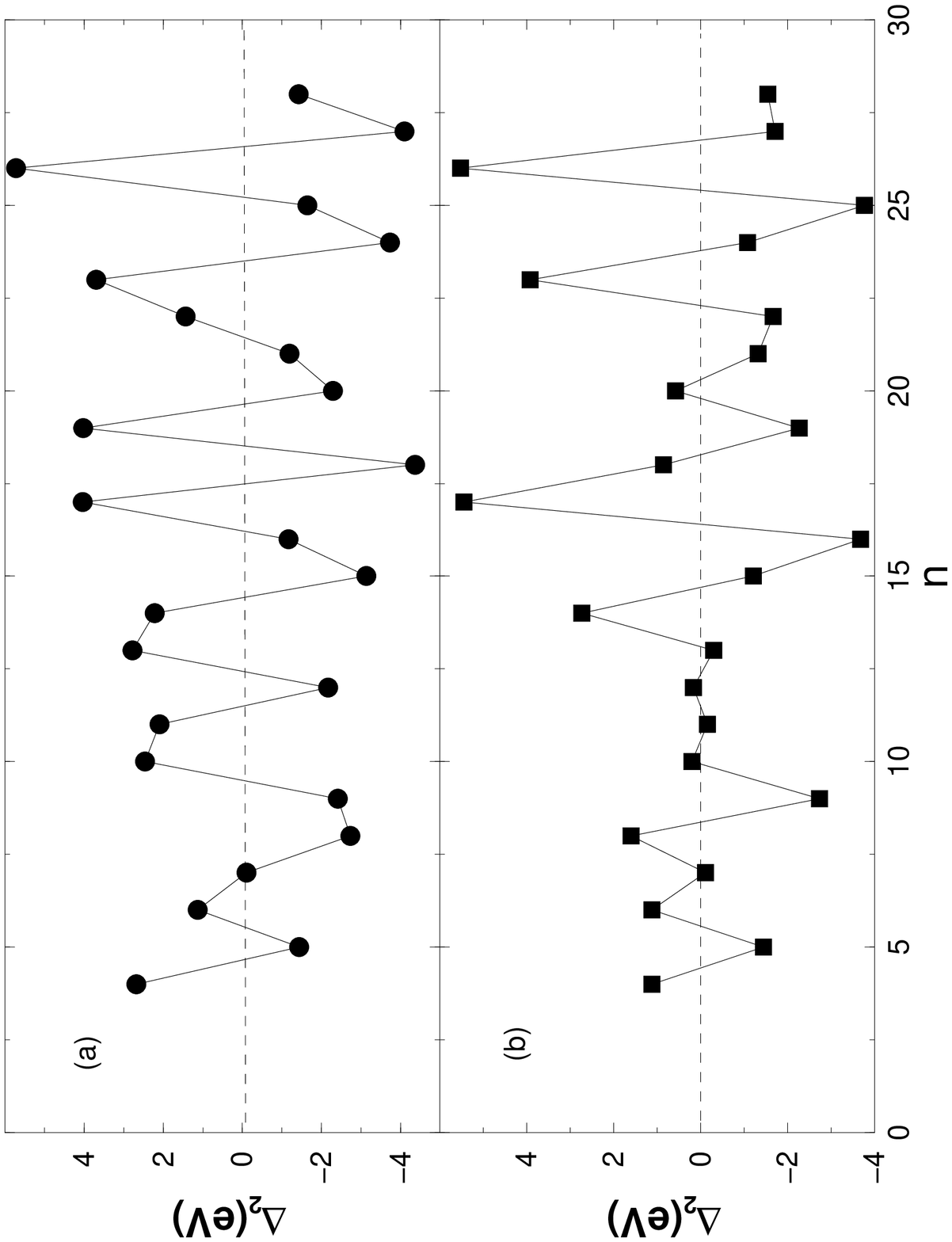}
\end{figure}

\end{document}